\magnification=1200
\pretolerance=10000
\baselineskip=24pt
\centerline{\bf Equilibrium configurations for quark-diquark stars}
\centerline{\bf and the problem of Her X-1 mass}

\bigskip
\centerline{J.E.Horvath$^1$ and J.A.de Freitas Pacheco$^2$}

\bigskip
\centerline{\it $^1$Instituto Astron\^omico e
 Geofisico, Av. Miguel Stefano 4200 - Agua Funda}
\centerline{\it 04301-904, S.Paulo, Brazil}

\centerline{\it $^2$ Observatoire de la C\^ote d'Azur, B.P.229, F-06304}
\centerline{\it Nice Cedex 4, France}

\bigskip
\noindent
{\bf Abstract}

We report new calculations of the physical properties of a quark-diquark
plasma. A vacuum contribution is taken into account and is responsible
for the appearance of a stable state at zero pressure and at a baryon
density of about 2.2 times the nuclear matter density in this
 model. The resulting
equation of state was used to integrate numerically the
Tolman-Oppenheimer-Volkoff equations.
The mass-radius relationship has been derived from a series of equilibrium
configurations constituted by a mixture of quarks and diquarks. These 
stellar
models, which are representative of a whole class,
 may be helpful to understand the possible compactness of the 
X-ray source Her X-1 and related objects. 

\vfill\eject

\noindent
{\bf 1.Introduction}

The main static properties of neutron stars can be studied in the mass-radius 
(hereafter $M-R$)
plane. As is well known, 
curves in such a diagram depend on the adopted equation of state (EOS).
Generally speaking, for a given EOS in which the matter does not 
have a self-bound state, the radius of an equilibrium 
configuration decreases
as the mass increases until a maximum (mass) is reached (see, for instance,
Ref. 1). Objects  less massive and
more compact  satisfying  the condition  $dM/dR \, > \, 0$ are unstable. 

From an observational point of view, the analysis of a given object in the
$M-R$ plane requires the determination of both quantities. Even though masses
can be obtained for neutron stars in binary systems, the radius is a
quantity of difficult direct access. Some constraints in the
M/R ratio have been established, for example, from the analysis of a sample of  
X-ray  bursters $^{2}$.

More recently, Li, Dai and 
Wang$^{3}$ have examined the location of the binary X-ray
pulsar Her X-1 in the plane $M-R$ plane. The mass ($0.98 \pm 0.12 M_{\odot}$)
 was calculated using
the orbital parameters of the system, whereas the radius of
the pulsar ($6.7 \pm 1.2$ km)
 was estimated using the
accretion torque model$^{4,5}$.
The compactness of Her X-1 was previously considered by
M\'esz\'aros and Riffert$^{6}$, who obtained an even smaller value for
the radius (5-6 km) using similar arguments.
 In spite of the indirect nature of this 
estimate, it is seems interesting to investigate the
consequences when these observational quantities are compared with
theoretical expectations. In fact, as Li,Dai and Wang 
noticed, the majority of
the existing  EOS predict a considerably higher radius, for
a stable configuration of that mass. In particular,  pure hadron stars 
(see Ref.1) or
models including a quark core, (see Ref. 7 and references therein) 
predict a radius of about 11.5 km  for a star
 with the same mass as Her X-1. Thus, if the
accretion torque and the cyclotron line formation theories are correct,
there is an indication that Her X-1 could be more compact than the 
configurations expected from models based on the present known EOS. 

In their recent work Li, Dai and Wang further 
suggested that Her X-1 could be a strange star, since these
objects could be more compact than neutron stars for certain 
parametrised QCD quantities, and thus 
more consistent with these derived properties.
The stability of strange matter and strange star models using a simple
EOS were considered first by Witten$^{8}$ 
and Farhi and Jaffe$^{9}$.
More detailed models and their physical properties were investigated
by Haensel et al.$^{10}$ and Alcock et al.$^{11}$, whereas their mechanical
properties were explored by Benvenuto and Horvath$^{12}$. However, as 
pointed out by Madsen$^{13}$ the value of the most important QCD parameter 
(namely the vacuum energy $B_{bag}$) needed for a compact strange star is not 
consistent with the postulated self-boundness of this phase. Therefore, 
strange matter is not likely to provide an acceptable model for 
Her X-1.

Generally speaking, 
a plasma of quarks and gluons is expected to form at some level 
as a consequence of
the disassembling of neutrons in stellar matter at densities exceeding
that of the nuclear matter (0.16 $fm^{-3}$). However, it
may be possible that the deconfinement of hadrons occurs through an
intermediate stage, in which nucleons are dissociated but quarks are
correlated in spin-singlet pairs, known as diquarks (for a review of 
diquark properties, see Ref.14). Since the suggestion
by Donoghue and Sateesh$^{15}$ that neutron matter may dissolve into
a {\it highly correlated quark fluid}, several papers were devoted
to discuss the physical properties of such a state. Horvath et 
al. (Ref. 16, hereafter paper I) calculated the properties of a diquark-quark
mixture, in which diquarks interact through a boson field described by
a "${\lambda \phi}^{4}$ "effective potential. Kastor and Traschen$^{17}$ have
previously computed a series of stellar models and their corresponding
$M-R$ relationship. However, because of the inclusion of an external
low-density neutron envelope below a threshold density, their models
are not suitable to explain the
compactness of Her X-1 since they behave akin to the hybrid stars$^{7}$.

In the present work we revisited the EOS of a diquark-quark mixture, including
the vacuum contribution. The vacuum energy density introduces
a negative pressure able to produce an absolute stable state$^{18}$, 
and open the possibility for the existence of a self-bound 
pure quark-diquark
star. We report in this paper calculations of 
relativistic star configurations and discuss how they relate to the
Her X-1 compactness problem.

\bigskip
\noindent
{\bf 2. The Equation of State}

\noindent
{2.1.Diquarks}

In order to calculate the EOS for a quark-diquark mixture, we
consider the spin-dependent colormagnetic interaction leading
to a spin-0 state, whose Hamiltonian is given in paper I.
Diquarks of mass $m_D$ are assumed to interact through 
a $\lambda\phi ^4$ effective potential due 
to a bosonic field, and precludes a Bose condensate.
The coupling parameter $\lambda$ may be estimated from
the $\Delta$-N mass difference and a variant of the P-matrix
formalism$^{19}$. Since the coupling constant $\lambda$
is found to be several orders of magnitude larger than the
critical value $\lambda_{*} = \, {{4\pi m_{D}^{2}} \over {m_{Planck}^{2}}} \,
\approx 3 \times 10^{-36}$, we are led to use the results of
Colpi et al. $^{20}$, who obtained parametric solutions for the
EOS of particles interacting through such an effective potential.
In this case, the pressure and the energy density due to
diquark-diquark interactions are respectively
$$ P_D = \varepsilon_0 (z-1)^2 \eqno(1)$$

and
$$\varepsilon_D = \varepsilon_0 (z-1)(3z+1) \eqno(2)$$

where $z \geq 1$ is a dimensionless parameter and $\varepsilon_0$
is an energy-density scale. If we adopt the same values as in paper I, namely 
 $m_{D} \,  = \, 575 \, Mev$ for the diquark mass 
and 
$\lambda \, = \, 27.8$ for the coupling constant, 
the energy density scale is found to be 
$\varepsilon_0 = 32.1 \, Mev \, fm^{-3}$.

\noindent
{2.2.Quarks}

Due to the broken chiral symmetry just after deconfinement, quarks acquire
a mass near the constituent  value, i.e., $m_q \approx 360 MeV$, which we
assume to be the same for both flavors $\it{u, d}$. If we further assume
that the free quarks constitute a Fermi fluid, the energy density 
for a given flavor is
$$
\varepsilon_{u, d} = {{3\over 10}{{\pi^{4/3}\hbar^2}\over m_q}n_{u, d}^{5/3}}
\eqno(3)$$

while the pressure is given by
$$ P_{u, d} = {{1\over 5}{{\pi^{4/3}\hbar^2}\over m_q}n_{u, d}^{5/3}} \eqno(4)$$

where $n_{u, d}$ stands for the number density of quarks u or d .

\noindent
{2.3.The quark-diquark mixture}

The properties of the quark-diquark plasma are calculated neglecting the
contribution from a leptonic component (electrons and muons) because 
they are not needed to neutralize the bulk matter.
In this case, the first condition to be satisfied is the electrical charge
neutrality, i.e.,
$${1\over 3}n_D + {2\over 3}n_u -{1\over 3}n_d = 0 \eqno(5)$$

The second condition states the conservation of the baryon charge, namely,
$$n_B = {2\over 3}n_D + {1\over 3}(n_u + n_d) \eqno(6)$$

where $n_B$ is the total baryon number density.

The total pressure of the mixture is
$$P = P_D + P_u + P_d - B \eqno(7)$$

where $B$ is the energy density of the vacuum, which in our
computations it is assumed to be equal to $57 \, MeV \,fm^{-3}$. We note 
that this choice is made without assuming that $B$ must be the same 
as the original MIT bag, since there are no compelling reason to force 
such an identification (see below).
Finally, the total energy density is given by

$$\varepsilon =  m_q c^2 (n_u + n_d) +  \varepsilon_D + 
\varepsilon_u + \varepsilon_d + B$$

where the first term on the right represents the rest energy density
of the quarks.

In order to obtain the equilibrium number density of
each constituent, the total energy density
was minimized with respect to $n_D$, keeping $n_B$ constant. In fact,
the condition $(\partial\varepsilon/\partial n_D)_{n_B}$ = 0 is
equivalent to impose the equality between
the chemical potential of diquarks and the sum of the chemical potential
of the two quark flavors. 

The system of equations above was solved numerically. For a given value
of the total baryon density $n_B$, we have computed the number density and
the pressure contribution of each constituent, as well
 as the total pressure and
the total energy density. The results of our calculations are given in 
Table 1. In
the first four columns, we give respectively (in $fm^{-3}$) the number density
of baryons, diquarks and of the $u$ and $d$ quark flavors. 
In the three following columns, we
give the pressure (in $GeV \, fm^{- 3}$) contributed by diquarks, 
by the sum of $u$ and $d$ flavours ($P_{u+d}$) and the
total pressure, 
including the vacuum contribution. The last column gives the total energy 
density of the mixture (in $GeV \, fm^{-3}$). 
From these results we note that diquarks carry, on
the average, about 60\% of the baryonic charge. At high densities, diquarks
contribute to about a half of the total pressure, and that fraction
increases as 
the density decreases. Most of the $u$ quarks are paired with $d$'s and thus
their contribution to the pressure and energy density is quite small. For
baryon densities smaller than 0.6 $fm^{-3}$, all $u$'s 
are paired and the diquark
number density is equal to that of $d$ quarks. A similar conclusion was
also reached by Kastor and Traschen$^{17}$. The sound velocity in such a
fluid can be obtained from the relation
$v_s = c(\partial P/\partial \varepsilon)^{1/2}$. From our results,
${v_s\over c} \, \approx \, 0.55$, indicating no violation of causality.
An important
point to be emphasized
is that at a baryon density of 0.36 $fm^{-3}$ (about 2.2 times
the nuclear density),
the total pressure of the mixture 
is zero, since the adopted vacuum pressure compensates
the contribution due to $d$ quarks and diquarks. In these conditions, our
calculations indicate that the energy per baryon number unit 
is about 834 $MeV$, smaller
than the nucleon rest energy. This corresponds to a stable state (see
Ref. 18 for a discussion)
allowing the existence of a pure quark-diquark star. We note also that
the energy density at $P \, = \, 0$ is about $7 \, B$, 
whereas for a pure quark plasma
such a value for the energy density is $4 \, B$. A plot of the equation of
 state is given in Fig.1.

\noindent{\bf 3.Stellar Models}

The equation of state obtained for the quark-diquark mixture and
the well known Tolman-Oppenheimer-Volkoff (TOV) equations were integrated
numerically, following the same procedure as Freitas Pacheco et
al.$^{7}$. For a given baryonic central density, the TOV equations
were integrated outwards until the condition $P \, = \, 0$ is satisfied.

The resulting $M-R$ relationship obtained from these calculations is
shown in Fig.2, while the plot of $n_{B} - M$ is shown
in Figure 3.  Stars of low mass ($M< 0.8 M_{\odot}$) are nearly uniform,
since the density ratio between the center and the surface is less than 
a factor of two. 
The maximum mass for a stable configuration is 1.18 $M_{\odot}$,
 corresponding to a radius of 7.35 km and a central density of
1.1 $fm^{-3}$, or about 6.6 times the density of nuclear matter. In
comparison with pure quark stars, diquark-quark stars
are more compact, and this may be partially attributed to the
role played by the (poorly known) confinement forces represented by 
the vacuum constant $B$ and the very presence of diquarks.

It is interesting to 
compare with the $M-R$ relation for strange stars$^{8,10,11,12}$ 
for the same value of the 
vacuum constant 
57 $MeV \, fm^{- 3}$. Those strange star models 
are not able to provide
the $M-R$ values needed for Her X-1. In fact, Li et al. obtain a 
compatibility only for strange stars models in which the
vacuum energy density is in the range 120-200 $MeV \, fm^{-3}$, considerably
higher than the MIT scale and certainly 
inconsistent with the strange matter hypothesis$^{9}$. 
It is only by imposing those high values of $B$ that the
required compactness is obtained$^{13}$. 
In our models the quark-diquark mixture produces
a softer EOS than a pure quark plasma and, as a consequence, a more
compact equilibrium configuration for the same mass and vacuum energy
density.

If Her X-1 is really quite compact, than a stable pure quark-diquark star is
compatible with the required $M - R$ values, without forcing the acceptable
range of physical parameters like the vacuum energy density or the
particle effective masses. However the maximum stable mass for these
particular models is slightly 
lower than the derived value of the more massive member in the binary
pulsar PSR 1913+16 (1.44 $M_{\odot}$). It's worth
mentioning that other X-ray sources other than Her X-1 are
suspect of being very compact. If the 4.1 keV line observed
in the  burster MXB 1636-536 is a gravitationally redshifted $K\alpha$
 $Fe$ line$^{21}$, then the use of a "canonical" neutron
star forces a radius of only 6.7 km. However,
the inclusion of the transverse Doppler effect due to  rotation
of the accreted gas$^{22}$ may increase the radius up to
10 km, compatible with neutron stars modeled through some
popular EOS. Fujimoto and Gottwald$^{23}$ based on the
assumption that the burster MXB 1728-34 attained the
Eddington limit at maximum flux, and modeling the X-ray
" color " evolution, concluded that the compact object
must have a mass of $M \approx 1.2 M_{\odot}$ and a 
radius of $R \approx 7.5 km$. These values match
those we derived for a diquark-quark star
near the instability limit. It is important to note that at least 
three independent methods have been employed to determine these 
low-radius figures.

\noindent
{\bf 4.Conclusions}

We have reported new calculations of the physical conditions
prevailing in a mixture of  deconfined quarks and diquarks,
satisfying the condition of zero charge. The inclusion
of confining forces (vacuum) induces the existence of a stable
state of the mixture 
at $P \, = \, 0$, and at a baryon density of about 2.2 times the
nuclear matter density. This state has an energy density of
about $7 \, B$, which should be compared to the 
value $4 \, B$ derived for stable strange matter$^{8}$. 
Moreover, it is clear that this property is shared by a family 
of EOS of the same type parametrised by the vacuum energy $B$. 
Although we have only given our results for a fixed $B$ numerically 
equal to the MIT bag constant, this quantity should be considered 
as a parameter 
measuring of the uncertain behaviour of the QCD vacuum and need 
{\it not} be related to the former. 

The TOV equations were solved numerically, and the resulting
equilibrium configurations are more compact than strange
stars. In particular, a quark-diquark star is able to explain
the $M-R$ values of Her X-1, if we accept that the radius is
correctly estimated from the accretion torque theory.

The softness of the resulting EOS for the quark-diquark
mixture implies in a maximum stable mass of only
1.18 $M_{\odot}$, and thus the presented models would not be 
able to explain
the nature of other compact stars like the more massive
component of the binary pulsar PSR 1913+16. 
We have emphasized that there is some evidence that 
other objects like MXB 1728-34 may
also be quite compact. In spite of the fact that the mass
and the radius are estimated from simple models,
(and therefore subject to some change) 
the compactness of these X-ray sources certainly raises
some problems for which these exotic models may be useful. 
The main reason is that $M \, \rightarrow \, 0$ when 
$R \, \rightarrow \, 0$ for stars of self-bound matter 
and the models are thus "naturally" compact for relatively 
low masses. 
It is clear that if the orthodox interpretation in terms
of nothing but one type of neutron star is maintained, 
then a softer
EOS is required in order to produce the necessary
compactness, but probably we have to face the problem of
a theoretical reduction of 
the maximum mass for the last stable object and conflict with 
the binary pulsar mass determination. 
The main point made in our work has been to show the existence of 
a whole class of exotic models for which we state that a 
subset of them$^{24}$ can 
satisfy both the binary pulsar mass bound and the compactness of 
Her X-1 and related sources.
Alternatively, the evidence 
could be taken as a hint for the existence of at least two types of 
compact stars, but if this is the case there would be no compelling 
need to accomodate both classes in a single stellar sequence. 

\vfill\eject

\centerline{\bf Table 1}

$$\vbox {\settabs 8\columns
\+$n_B$&$n_D$&$n_u$&$n_d$&$P_D$&$P_{(u+d)}$&$P$&$\varepsilon$& \cr
\+$fm^{-3}$&$fm^{-3}$&$fm^{-3}$&$fm^{-3}$&$GeV fm^{-3}$&
$GeV fm^{-3}$&$GeV fm^{-3}$&$GeV fm^{-3}$& \cr

\+0.36&0.36&0.00&0.36&0.039&0.018&0.000&0.40& \cr
\+0.40&0.40&0.00&0.40&0.046&0.021&0.010&0.45& \cr
\+0.48&0.48&0.00&0.48&0.063&0.030&0.036&0.55& \cr
\+0.60&0.57&0.03&0.63&0.082&0.047&0.072&0.71& \cr
\+0.84&0.77&0.07&0.91&0.128&0.086&0.157&1.05& \cr
\+1.16&1.04&0.12&1.29&0.200&0.154&0.298&1.56& \cr
\+1.51&1.34&0.17&1.68&0.289&0.243&0.475&2.15& \cr
\+1.88&1.66&0.22&2.10&0.393&0.351&0.687&2.82& \cr} $$

\noindent
Table 1 caption: Physical properties of an equilibrium 
quark-diquark mixture. Units appear as indicated, see text for details and 
Fig.1.

\vfill\eject

\centerline {\bf Figure captions}
\vskip 1 true cm
Figure 1. Equation of state of the quark-diquark matter (solid line). The 
partial pressures contributed by the diquarks $P_{D}$ (long dashed line) 
and quarks $P_{u+d}$ (short dased line) are also shown. 

Figure 2.$M-R$ plot for quark-diquark stars. 
Her X-1 mass limits are shown with dashed lines. The models indicate 
a radius of $\sim \, 7$ km in agreement with the the analysis by
Li et al. (Ref.3).

\bigskip
Figure 3. $n_{B}-M$ plot for quark-diquark stars. For the Her X-1 mass 
renge the models have a central density of about six times the nuclear 
matter value.

\vfill\eject
\noindent
{\bf 5.References}

\noindent
1) D.Arnett and R.L.Bowers, {\it Astrophys.J.Supp.}{\bf 33}, 415 (1977).

\noindent
2) H.Inoue, in {\it The Origin and Evolution of Neutron Stars, 
IAU Symposium  125}, eds. D.J.Helfand and J.-H. Huang (Reidel, Dodretch,1986) 
p.233.

\noindent
3) X.-D.Li, Z.-G.Lai and Z.-R.Wang, {\it Astro.Astrophys.}{\bf 303}, L1 (1995).

\noindent
4) I.Wasserman and S.L.Shapiro, {\it Astrophys.J.}{\bf 265}, 1036 (1983).

\noindent
5) X.-D.Li and Z.-R.Wang, {\it Astrophys.Space Sci.}{\bf 225}, 47 (1995) ; 
X.D.Li and Z.R.Wang, {\it Astrophys.Space Sci.}{\bf 234}, 39 (1995).

\noindent
6) P.M\'ez\'aros and H.Riffert, {\it Astrophys.J.}{\bf 323}, L127 (1987).

\noindent
7) J.A.de Freitas Pacheco, J.E.Horvath, J.C.N.de Ara\'ujo and M.Cattani, 
{\it Mon.Not.R.A.S.}{\bf 260}, 499 (1993).

\noindent
8) E.Witten, {\it Phys.Rev.D}{\bf 30}, 272 (1984).

\noindent
9) E.Farhi and R.L.Jaffe, {\it Phys.Rev.D}{\it 30}, 2379 (1984).

\noindent
10) P.Haensel, J.L.Zdunik and R.Schaeffer, {\it Astron.Astrophys.}{\bf 160}, 
121 (1986).

\noindent
11) C.Alcock, E.Farhi and A.V.Olinto, {\it Astrophys.J.}{\bf 310}, 261 (1986).

\noindent
12) O.G.Benvenuto and J.E.Horvath, {\it Mon.Not.R.A.S.}{\bf 250}, 679 (1991).

\noindent
13) J.Madsen, astro-ph 9601129 preprint (1996).

\noindent
14) M.Anselmino, E.Predazzi, S.Ekelin, S.Fredriksson and 
D.B.Lichtenberg, {\it Rev.Mod.Phys.}{\bf 65}, 1199 (1993).

\noindent
15) J.Donoghue and K.S.Sateesh, {\it Phys.Rev.D}{\bf 38}, 360 (1989).

\noindent
16) J.E.Horvath, J.A.de Freitas Pacheco and J.C.N.de Ara\'ujo, {\it Phys.Rev.D}
{\bf 46}, 4754 (1992), paper I.

\noindent
17) D.Kastor and J.Traschen, {\it Phys.Rev.D}{\bf 44}, 3791 (1991).

\noindent
18) J.E.Horvath, {\it Phys.Lett.B}{\bf 294}, 412 (1992).

\noindent
19) R.L.Jaffe and F.E.Low, {\it Phys.Rev.D}{\bf 19}, 2105 (1979).

\noindent
20) M.Colpi, S.L.Shapiro and I.Wasserman, {\it Phys.Rev.Lett.}{\bf 57}, 
2485 (1986).

\noindent
21) I.Waki et al., {\it PASJ}{\bf 36}, 819 (1984).

\noindent
22) M.Y.Fujimoto, {\it Astrophys.J.}{\it 293}, L13 (1985).

\noindent
23) M.Y.Fujimoto and M.Gottwald, {\it Mon.Not.R.A.S}{\bf 236}, 545 (1989).

\noindent
24) J.E.Horvath and J.A.de Freitas Pacheco, in preparation.

\bye